\documentclass[11pt]{article}
\usepackage{wds11,epsf}

\begin{document}

\title{The Angular Correlation Function of Quasars from SDSS DR3}

\author{G.Yu. Ivashchenko, V.I. Zhdanov}

\affil{Kyiv National Taras Shevchenko University, Kyiv, Ukraine.}

\begin{abstract}
We estimate the two-point angular correlation function (CF) of
quasars from SDSS DR3 using a special method of comparison random
catalog generation. The best-fit value for the CF power-law index
is found to be $\alpha=0.78\pm0.18$ on the $2<\theta<250$ $arcmin$
interval. This is lower (though in marginal agreement) than
earlier result of \markcite{{\it Myers et al.} (2005)} based on
SDSS DR1 catalogue of photometrically-classified quasars.
\end{abstract}

\begin{article}

\section{Introduction}

Two-point correlation functions (CF) of extragalactic objects give
us effective means for investigating the large scale structure
(LSS) of the Universe. Most complete information can be obtained
from the real-space three-dimensional CF, which is convenient for
verification of the LSS models. Estimation of this CF is based on
linear distances and involves additional (cosmological)
parameters. On the other hand, the angular CF is most directly
connected with observational data and its estimation practically
does not require additional information about the cosmological
model. The source of data for CF estimation are extragalactic
surveys. When studying the distribution of matter at high
redshifts, investigation of quasars is of particular interest,
because they are very luminous. In the present paper, we consider
the angular two-point correlation function of these objects. We
employ the third edition of the SDSS Quasar Catalog
(\markcite{{\it Schneider et al.} 2005}), which consists of the
46,420 objects brighter than those with absolute magnitudes
$M_{i}=-22$. The area covered by the catalog is $\approx 4188\
deg^{2}$. The quasar redshifts range from 0.08 to 5.41, however in
this paper we restrict ourselves to $z<2.4$ . Note that the
two-point correlation function of objects from the first edition
of the SDSS catalogue of photometrically-classified quasars has
been estimated by \markcite{{\it Myers et al.} (2005)}. Here we
use the SDSS third data release, which is the result of a more
reliable classification method (see \markcite{{\it Schneider et
al.} (2005)} and references therein).

The surveys of quasars have large redshift depth, but they include
fewer objects than the galaxy surveys. This complicates the CF
estimation. Furthermore the surveys have their own selection
effects, in particular, because of their finite volume,
heterogeneity or receiver properties, etc. The essential stage of
the CF estimation is the construction of a random catalog taking
into account these selection effects as much as possible, which
allows to determine the excess of object pairs over random
background. There are different methods for such constructions
that allow to take into account the initial catalogue properties,
and it is useful to verify the results by different statistical
means. In the present paper, we use the methodology of artificial
catalog generation by \markcite{{\it Zhdanov \& Surdej} (2001)},
which retains some selection effects of the initial catalogue due
to the heterogeneity of the spatial sample.

\section{Angular Correlation Function}

According to \markcite{{\it Peebles } (1980)}, the angular
two-point correlation function of the object distribution
$\omega(\theta)$ determines the probability to find two objects
with positions inside small areas $d\Omega_{1}$ and $d\Omega_{2}$
on the unit sphere
\begin{equation}\label{dP}
    dP=n_{0}^{2}[1+\omega(\theta)]d\Omega_{1}d\Omega_{2},
\end{equation}
where $n_{0}$ is the average object density on the celestial
sphere in the area involved. The total number of pairs having
separation $\theta \in d\Omega$ is
$dN_{p}=\frac{1}{2}N_{t}n_{0}d\Omega[1+\omega(\theta)]$, $N_t$
being the total number of objects in the sample. If there are no
pair correlations in a (random) catalogue, the number of pairs
equals to $dN_{p}^{\ast}=\frac{1}{2}N_{t}n_{0}d\Omega$. The excess
of the pair number $dN_{p}$ over the random background
$dN_{p}^{\ast}$ is
\begin{equation}\label{dN}
    dN_{p}-dN_{p}^{\ast}=\frac{1}{2}N_{t}n_{0}\omega(\theta)d\Omega.
\end{equation}
Estimation of the random background is very important, because it
is specific in different catalogues.

It is convenient to use the average number density of pairs from
the given range of $\theta^{'}$ and $\Delta z$, calculated from
the initial catalogue:

\begin{equation}\label{n}
   n(\theta,\xi)=
   \frac{N(\theta\leq\theta^{'}\leq\theta+\beta,\xi-\zeta\leq\Delta
   z\leq\xi+\zeta)}{2\pi\beta(\theta+\frac{\beta}{2})},
\end{equation}
 where $\xi$ and $\beta$ define the discretization interval for
 the redshift z and angle $\theta$. Also we introduce
 the similar value $n^{\ast}(\theta,\xi)$ for the randomized
catalogue,.

Let $\xi_{c}$ be a redshift correlation length. Calculation of the
excess number $N_{p}-N_{p}^{\ast}$ in fact involves only the pairs
with $|\Delta z|<\xi_{c}$, when we expect a physical relationship
between the objects. Therefore, in order to estimate the two-point
angular CF according to (\ref{dN}), we use the following relation:
\begin{equation}\label{cf}
    \omega(\theta)=\frac{2[n(\theta,0)-n^{\ast}(\theta,0)]}{n_{0}N_{t}},
\end{equation}
where we set $\zeta=0.1$, i.e. we include only quasar pairs with
$\Delta z<0.1$. Typically we used discretization bin size
$\Delta\theta=60''$ and $\Delta\theta=100''$.

Random catalogues were generated using the method of
\markcite{{\it Zhdanov \& Surdej} (2001)}. The values of redshift
were randomly rearranged in the initial catalogue list of objects,
while the values of the right ascension and declination were left
in the same order. Hence the real physical correlations of pairs
were washed out, but the angular distribution of the quasars has
been preserved. In this way we generated 100 randomized catalogues
and the value of $n^{\ast}(\theta,0)$ was obtained as the mean
value.

To verify the results we also calculated the pair number densities
$n(\theta, \xi)$ for several values of $\xi>0.3$ and the mean
value $‹n(\theta,\xi)›_{\xi}$ over $\xi$. For these pairs, no
physical correlations are expected. We note the satisfactory
coincidence of $n^{\ast}(\theta,0)$ and $‹n(\theta,\xi)›_{\xi}$
 within the wide range of angle values; this testifies the good quality
 of the randomized catalogues.

\section{Results}

The values of the two-point angular correlation function (Fig. 1),
defined by formula (\ref{cf}), were approximated by the dependence

\begin{equation}\label{cf_th}
    \omega(\theta)=a+b\theta^{-\alpha},
\end{equation}
where  $\alpha$ is the index of the correlation function; the
 constant $a$ has been introduced to compensate possible
uncertainty of the background pair numbers, but it was found to be
statistically insignificant. The results were checked for
different regions of SDSS DR3. The best-fit values of $\alpha$ are
presented in Table 1 for the quasars with $0.3<z<2.4$. For
comparison the results from Fig. 5 of \markcite{{\it Myers et al.}
(2005)} are also presented.

\begin{figure}[htb]
\begin{center}
\begin{tabular}{c}
  \epsfxsize=130mm
  \epsfysize=100mm
  \epsfbox{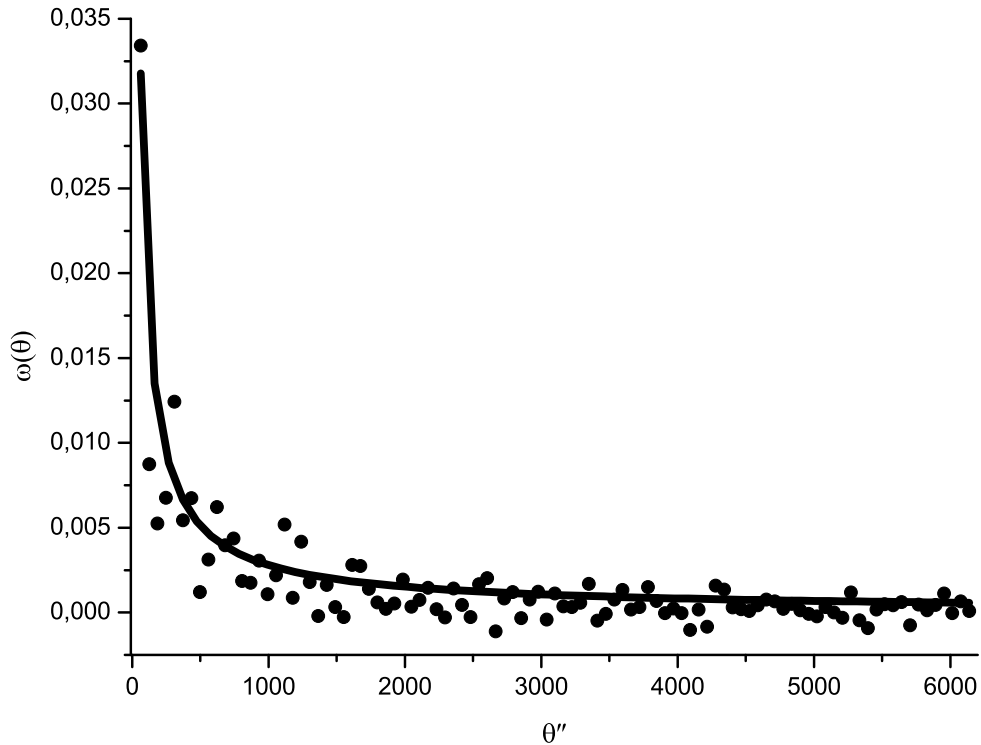}
\end{tabular}
\end{center}
\caption{Two-point angular correlation function $\omega(\theta)$
and its approximation (\ref{cf_th}) }
\end{figure}

\begin{table}
\caption{Index of the two-point angular correlation function}
  \centering
\label{tabl1}
\begin{tabular}{|c|c|c|}
  \hline
Source & $\alpha$ & Interval, arcmin \\
  \hline
  SDSS DR3 & $0.70\pm0.14$ & [1,250] \\
  SDSS DR3 & $0.84 \pm0.22$ & [5,250] \\
  SDSS DR3 & $1.05\pm0.20$ & [3,150] \\
  SDSS DR3 & $0.93\pm0.27$ & [5,150] \\
  SDSS DR3 & $0.78\pm0.18$ & [2,250] \\
  SDSS DR1 \markcite{{\it Myers et al.}(2005)} & $0.98\pm0.15$ & [2,250] \\
  \hline
\end{tabular}

\end{table}

\section{Discussion}

The result of fitting on the largest interval $1<\theta<250$
$arcmin$ is  $\alpha=0.70\pm0.14$. As one can see from Table 1,
our results are in marginal agreemrent with that of Fig.5 from
\markcite{{\it Myers et al.} (2005)} within $2\sigma$ errors. On
the other hand the parameter $\alpha$ in the case of quasars is
not found to be essentially larger than the usual value
$\alpha\approx0.7$ for galaxies, including more detailed galaxy
type analysis (e.g., \markcite{{\it Connolly et al.} 2002},
\markcite{{\it Budavary et al.} 2003}). Different values of
$\alpha$ on different scales may be due to poor statistics;
however we cannot cast aside a possibility that this is a result
of LSS peculiarities like cellular structures. Anyway the function
(5) may be considered only as an approximate model (cf., e.g.,
\markcite{{\it Connolly et al.} 2002}, \markcite{{\it Myers et
al.} 2005}). We tried to modify (5) for small angles, but this did
not give rise to any statistically significant results.

We are thankful to our referees for useful suggestions and
comments.

\end{article}

\end{document}